# A note on hypoplastic yielding

J. J. Nader

Department of Structural and Geotechnical Engineering, Polytechnic School,
University of São Paulo, Brazil

**SUMMARY**

This note discusses briefly the definition of yield surface in hypoplasticity in connection with the physical notion of yielding. The relation of yielding with the vanishing of the material time derivative of the stress tensor and the vanishing of the corotational stress rate is investigated.

KEY WORDS: hypoplasticity, yielding, yield surface, hypoelasticity.

## INTRODUCTION

This note deals with the connection between the physical notion of yielding and the way the yield surface is determined in hypoplasticity. An inspection of the pertinent literature shows that the present state of affairs is not completely clear.

Let us start the discussion with the general hypoplastic equation:

$$\overset{\circ}{\mathbf{T}} = \mathbf{h}(\mathbf{T},\mathbf{D}), \qquad (1)$$

where $\mathbf{T}$ is the Cauchy stress, $\overset{\circ}{\mathbf{T}}$ is the corotational stress rate, $\mathbf{D}$ is the stretching tensor, and $\mathbf{h}$ is an isotropic positively homogeneous function of degree one in $\mathbf{D}$ (see [1] for articles treating several aspects of this theory).

In hypoplasticity one obtains the yield surface equation by searching pairs $(\mathbf{T},\mathbf{D})$ that satisfy $\mathbf{h}(\mathbf{T},\mathbf{D})=\mathbf{0}$ (see next section ), as is done in hypoelasticity [2]. The first elements of these pairs, *i.e.*, the stresses, form the yield surface. In accordance with this definition and in view of Eq. (1), researchers working in this field usually say that in order to find the yield surface equation one must impose that the corotational stress rate be equal to zero ($\overset{\circ}{\mathbf{T}}=\mathbf{0}$). On the other hand, in conformity with the physical notion of yielding, we say that a material yields when it deforms under constant stress; in this sense yielding is associated with the vanishing of the material time derivative of $\mathbf{T}$ ($\dot{\mathbf{T}}=\mathbf{0}$). This ambiguous situation led Kolymbas and Herle [3] to write: "States at which the material deforms without further stress changes (i.e. where plastic flow occurs) are called limit states.[...].Referring to the condition 'stress remains constant', one should specify whether $\dot{\mathbf{T}} = \mathbf{0}$ or $\overset{\circ}{\mathbf{T}} = \mathbf{0}$ is meant".

The objective of this brief note is to clear the matter. With this purpose Eq. (1) will be treated as a differential equation in the unknown function **T**, so the occurrence of yielding can be investigated in a time interval rather than at a single instant, as implied in the expression "stress *remains* constant". This is different from the usual approach of employing only algebraic means to study yielding.

The notation adopted here is standard in contemporary continuum mechanics.

## YIELDING AND THE YIELD SURFACE

Consider the set C of all ordered pairs (**T**, **D**), with **D**≠**0**, such that **h**(**T**, **D**)=**0**. The set of all first elements **T** of the pairs of tensors in C is the yield surface S. Due to the isotropy of **h**, for any **Q**∈ *Orth*, (**T**, **D**)∈ C if and only if (**QTQ**$^T$, **QDQ**$^T$)∈ C; and, therefore, **T**∈ S if and only if **QTQ**$^T$∈ S. In addition, since **h** is positively homogeneous of degree one in **D**, if (**T**, **D**)∈ C, then (**T**, λ**D**)∈ C, for any real number λ>0. For the representation of critical states it is further required that the tensors **D** present in the pairs of the set C are traceless, but this is not relevant for the purposes of this paper.

We will now discuss the connection between yielding and the definition of S given above. Firstly, it is convenient to introduce $\overset{\circ}{\mathbf{T}} = \dot{\mathbf{T}} - \mathbf{WT} + \mathbf{TW}$ in Eq. (1) to obtain:

$$\dot{\mathbf{T}} = \mathbf{h}(\mathbf{T},\mathbf{D}) + \mathbf{WT} - \mathbf{TW} , \qquad (2)$$

where **W** is the spin tensor. It is interesting to note in passing that S can be seen as formed by equilibrium points of Eq. (2) in certain motions [4], [5].

The results presented below (propositions A and B) are related to the following initial value problem, from now on referred to as problem H: given **D**:*I→Sym*, **W**:*I→Skw*, continuous functions of time *t* (*I* is an open interval of real numbers containing 0), find the solution **T**(*t*) of the differential equation (2), with the initial condition **T**(0)=**T**$_0$. We assume that the function **h** has smoothness properties that ensure existence and uniqueness of solution of problem H.

**A)** Let (**T**\*,**D**\*) belong to C. For any continuously differentiable **Q**:*I→Orth*, with **Q**(0)=**1** (**1** is the second-order identity tensor), consider problem H with **D**(*t*)=**Q**(*t*)**D**\***Q**$^T$(*t*), $\mathbf{W}(t) = \dot{\mathbf{Q}}(t)\mathbf{Q}^T(t)$ and **T**$_0$=**T**\*. Then the solution of problem H is

$\mathbf{T}(t) = \mathbf{Q}(t)\mathbf{T}^*\mathbf{Q}^T(t)$, and so $\overset{\circ}{\mathbf{T}}(t) = \mathbf{0}$ for all $t$ in $I$.

The proof is simple; it involves the isotropy of **h**.

Thus, to each function **Q** in (A) there corresponds a motion in which $\overset{\circ}{\mathbf{T}}(t)$ is permanently zero and therefore $\mathbf{T}(t)$ remains on S. In particular, if we choose for **Q** the constant function with value $\mathbf{Q}(t)=\mathbf{1}$, the solution of problem H is the constant function with value $\mathbf{T}(t)=\mathbf{T}^*$. This result means that it is possible to deform with zero spin a material whose initial stress is on S in such a way that the stress does not change, that is, in such a way that the material yields. But, since the motions considered in (A), each one produced by a choice of **Q**, differ from each other simply by a change in observer [6], it is reasonable to say that yielding occurs in all of them.

Note that, as **h** is positively homogeneous of degree one in **D**, (A) remains valid if we put $\alpha(t)\mathbf{D}^*$, with $\alpha(t)>0$, instead of the constant stretching $\mathbf{D}^*$.

It is important to remark, on the other hand, that, depending on the function **h**, there may be motions in which the stress point remains on S but no yielding occurs. The next result is an example of this fact.

(**B**) Let $(\mathbf{T}^*,\mathbf{D}^*)\in C$, with $\mathbf{T}^*\neq\mathbf{0}$, and assume that: 1) **h** is homogeneous of degree one in **T**; 2) there is a non-zero real $\beta$ and a symmetric tensor $\mathbf{D}^+$ such that $\mathbf{h}(\mathbf{T}^*,\mathbf{D}^+)=\beta\mathbf{T}^*$ (and hence, although $\mathbf{T}^*\in S$, $(\mathbf{T}^*,\mathbf{D}^+)\notin C$). Then the solution of problem H, with $\mathbf{D}(t)=\mathbf{D}^+$, $\mathbf{W}(t)=\mathbf{0}$ and $\mathbf{T}_0=\mathbf{T}^*$, is $\mathbf{T}(t)=\exp(\beta t)\mathbf{T}^*$, whose corotational rate, $\overset{\circ}{\mathbf{T}}(t)=\beta\exp(\beta t)\mathbf{T}^*$, does not vanish. But, since $\mathbf{h}(\mathbf{T}(t),\mathbf{D}^*)=\exp(\beta t)\mathbf{h}(\mathbf{T}^*,\mathbf{D}^*)=\mathbf{0}$, $\mathbf{T}(t)\in S$ for all $t$ in $I$.

The proof is simple too.

Hypotheses 1 and 2 in (B) are satisfied in many hypoplastic models, *e.g.*, [7], [1], [4]. In the CLoE model [8], however, hypothesis 2 is not satisfied.

We saw in (A) how to produce motions with $\overset{\circ}{\mathbf{T}}(t)=\mathbf{0}$. We could now ask: does the stress change necessarily as in (A) whenever $\overset{\circ}{\mathbf{T}}(t)=\mathbf{0}$ in a certain interval? The answer is in the affirmative, since the solution of $\dot{\mathbf{T}} = \mathbf{WT} - \mathbf{TW}$, with $\mathbf{T}(0)=\mathbf{T}_0$, is $\mathbf{T}(t)=\mathbf{Q}(t)\mathbf{T}_0\mathbf{Q}^T(t)$, $\mathbf{Q}(t)$ being the solution of $\dot{\mathbf{Q}} = \mathbf{WQ}$, with $\mathbf{Q}(0)=\mathbf{1}$. This conclusion is independent of the constitutive equation.

# FINAL REMARKS

The intention of this note was to review more precisely some statements about yielding. We have seen that, if the stress point is initially on the yield surface S, it is possible to deform the material in such a way that $\overset{\circ}{\mathbf{T}}(t) = \mathbf{0}$ in a time interval (and so the stress remains on S); in particular, if the spin is zero, the material can be deformed at constant stress ($\dot{\mathbf{T}}(t) = \mathbf{0}$). On the other hand, we have also seen that in many hypoplastic models the stress point can move on S with $\overset{\circ}{\mathbf{T}}(t) \neq \mathbf{0}$ (no yielding).

The results remain essentially the same in models that include the void ratio *e* as an argument of **h** (*e.g.*, [9], [10]). During yielding, as *tr***D**=0 (critical states), *e* remains constant.

Finally, it is interesting to observe that proposition A and the conclusions associated to it apply to hypoelasticity as well. Proposition B may apply to particular hypoelastic models.

*E-mail: jjnader@usp.br*

*Address: Av. Prof. Almeida Prado, Departamento de Engenharia de Estruturas e Geotécnica, Escola Politécnica, Universidade de São Paulo, 05508-900, São Paulo, Brasil.*